\documentclass[journal]{IEEEtran}

\usepackage{array}
\usepackage{amssymb}
\usepackage{amsthm}
\usepackage{amsmath}
\usepackage[printonlyused, nolist]{acronym}
\usepackage{steinmetz} 
\usepackage{multirow}
\usepackage{textcomp} 
\usepackage{gensymb}
\usepackage{color}
\usepackage{floatrow}
\usepackage[caption=false,font=small]{subfig}
\usepackage{graphicx}
\usepackage{epstopdf}
\usepackage{rotating}
\usepackage{bigstrut}
\usepackage[hyphens]{url}
\usepackage{units}
\usepackage[disable]{todonotes} 
\newcolumntype{C}[1]{>{\centering\let\newline\\\arraybackslash\hspace{0pt}}m{#1}}
\graphicspath{{./pics/}, {./case_study/pics/}}
\usepackage[hidelinks]{hyperref}

\begin{document}
\title{\texttt{pandapower} - an Open Source Python Tool for Convenient Modeling, Analysis and Optimization of Electric Power Systems}

\author{Leon~Thurner, Alexander~Scheidler, Florian~Sch\"afer, Jan-Hendrik~Menke, Julian~Dollichon, Friederike~Meier, Steffen~Meinecke and Martin~Braun~\IEEEmembership{Senior Member,~IEEE}
\thanks{Leon~Thurner,~Florian~Sch\"afer,~Jan-Hendrik~Menke,~Steffen~Meinecke~and~Martin~Braun are with the Department of Energy Management and Power System Operation at the University of Kassel, Germany}
\thanks{Alexander~Scheidler,~Julian~Dollichon,~Friederike~Meier~and~Martin~Braun are with the  Department of Distribution Systems Operation of Fraunhofer Institute for Energy Economics and Energy System Technology (IEE) in Kassel, Germany}}
\markboth{This paper has been accepted for publication in IEEE Transaction on Power Systems. \textcopyright 2018 IEEE.}%
{}

\maketitle

\begin{abstract}
\texttt{pandapower} is a Python based, BSD-licensed power system analysis tool aimed at automation of static and quasi-static analysis and optimization of balanced power systems. It provides power flow, optimal power flow, state estimation, topological graph searches and short circuit calculations according to IEC~60909. \texttt{pandapower} includes a Newton-Raphson power flow solver formerly based on \textsc{pypower}, which has been accelerated with just-in-time compilation. Additional enhancements to the solver include the capability to model constant current loads, grids with multiple reference nodes and a connectivity check. The \texttt{pandapower} network model is based on electric elements, such as lines, two and three-winding transformers or ideal switches. All elements can be defined with nameplate parameters and are internally processed with equivalent circuit models, which have been validated against industry standard software tools. The tabular data structure used to define networks is based on the Python library pandas, which allows comfortable handling of input and output parameters. The implementation in Python makes \texttt{pandapower} easy to use and allows comfortable extension with third-party libraries. \texttt{pandapower} has been successfully applied in several grid studies as well as for educational purposes. A comprehensive, publicly available case-study demonstrates a possible application of \texttt{pandapower} in an automated time series calculation.
\end{abstract}


\IEEEpeerreviewmaketitle

\section{Introduction}
\IEEEPARstart{A} paradigm shift in electric power systems towards distributed generation as well as an increasing degree of automation raises the complexity of power system operation, analysis and planning in future power systems. Challenges arise especially in distribution systems, where a majority of distributed energy resources are connected. The rising level of complexity calls for new tools that allow a high degree of automation, while still being easy to use. Open source tools provide a free and transparent alternative to commercial tools in scientific applications as well as for educational purposes \cite{PFENNINGER201863, Milano2009}.

\subsection{Available Open Source Tools}
There are several open source power system calculation tools with different strengths and focuses available today \cite{Milano2009}. \textsc{matpower} \cite{matpower} is a widely used power system analysis tool that solves power flow and optimal power flow problems and also includes an optimal scheduling tool for market simulations \cite{MOST}. There are ports of the original \textsc{matlab} based code to other languages, most notably Pythons \textsc{pypower} \cite{pypower}. MatDyn \cite{matdyn} and pypower-dynamics \cite{pypower-dynamics} extend \textsc{matpower} and \textsc{pypower} respectively for dynamic analysis of electric power systems. Dynamic network evaluations and simulations are also provided by  \textsc{psat} \cite{psat}, which is based on \textsc{matlab}/Simulink. Dome \cite{dome} is a Python tool which was derived from \textsc{psat}, but is not available under an open source license. GridCal \cite{gridcal} includes power flow, time-series and short circuit calculation methods and comes with a comprehensive graphical user interface. The simulation and optimization library PyPSA \cite{pypsa} is aimed at time-series simulation of security-constrained linear optimal power flow and investment optimization. PowerGAMA \cite{powergama} and psst \cite{psst} are also aimed at market optimization in electric networks. OpenDSS is a Delphi based static simulation tool that allows a wide range of network analysis including unbalanced power flow calculations \cite{opendss}. GridLAB-D is an advanced tool for static simulations using agent based simulation models, based on C/C++ with integration of SCADA controls, metering and market models \cite{gridlab}.

\subsection{Python in Power Systems Analysis}
Many of the available open source tools are based on \textsc{matlab} \cite{matpower, matdyn, psat} or Delphi \cite{opendss}. Even though the tools themselves are open source, they depend on commercial platforms. While open alternatives, such as GNU Octave instead of \textsc{matlab}, sometimes exist, they cannot always guarantee the same functionality as their commercial equivalents. Tools which are based on commercial platforms can neither be freely used as stand-alone software nor easily extended with other libraries. Parallelization on computational clusters is also subject to specific license agreements. The fact that Delphi is not designed to run on Linux further limits the possibility to deploy computational clusters. A free alternative to commercial platforms is the programming language Python, which is available under an open source license. Python is a scripting language with a straight-forward and easy to learn syntax. Scientific libraries like numpy and scipy are internally implemented in C, so that mathematical analysis and data manipulation routines are carried out efficiently \cite{scipy}. Python has gained significant popularity for open source projects, especially in scientific applications. Since a large variety of libraries are freely available in Python, Python applications can be easily extended with third-party libraries. They can also be parallelized on computational clusters without license or compatibility constraints. Consequently, many recently developed tools for power system analysis are implemented in Python~\cite{Milano2009, PFENNINGER201863, pypower, gridcal, pypsa, powergama, psst}.

\subsection{Structure and Overview}
This paper introduces the new power systems analysis tool \texttt{pandapower}. A general introduction into the motivation for the development of pandapower as well as basic design choices with regard to grid modeling and data structure are given in Section \ref{sec:pandapower}. The  extensive library of electric elements, such as ZIP loads, lines, transformers or switches that comes with pandapower is discussed in Section \ref{sec:Elements}. The electric analysis methods of pandapower are discussed in Section \ref{sec:electric_analysis}: The \texttt{pandapower} power flow implementation is originally based on \textsc{pypower}, but has been extended with several features and accelerated with just-in-time compilation (see Section \ref{sec:PowerFlow}). The optimal power flow allows using the interior point solver provided by \textsc{pypower} with the \texttt{pandapower} element models (see Section \ref{sec:OPF}). \texttt{pandapower} includes an original implementation of a weighted least squares state estimation including bad data detection (see Section \ref{sec:StateEst}). It also includes an original implementation of a short circuit calculation in accordance with IEC 60909 (see Section \ref{sec:SC}). On top of the electric analysis functions, a module for topological searches allows graph analysis of electric networks using the NetworkX library (see Section \ref{sec:Top}). A comprehensive case study with a quasi-static time-series simulation in an active distribution grid is given in Section \ref{sec:case_study}. Finally, a summary and conclusion is given in Section \ref{sec:Conclusion}.

\section{pandapower} \label{sec:pandapower}

\subsection{Motivation}
Since many commonly used power system calculation tools are developed  for the North American power system layout, they are either focused on balanced  transmission system analysis (e.g. \textsc{matpower}, \textsc{pypower}) or three-phase distribution grid models (e.g. OpenDSS, Gridlab-D). However, power systems in Europe and other parts of the world are designed symmetrically up until the end consumer connection point in the low voltage level. Symmetric grid modeling and analysis is therefore routinely used to analyze distribution systems. But even though there is a very large need for symmetric distribution system analysis, there is no open source tool specifically focused on automated symmetric distribution system analysis.

Available tools which are well suited for automation are developed for transmission systems (e.g. \textsc{matpower}, \textsc{pypower}), and do not include the possibility to define elements with nameplate parameters (see Section \ref{sec:BBM}). Other tools, which do include element models, are focused on dynamic calculations (e.g. Dome, \textsc{PSAT}), energy optimization (e.g. PyPSA) or unbalanced analysis (e.g. OpenDSS, Gridlab-D). Additionally, some functionality such as short circuit calculations according to IEC~60909 or graph searches are not available in the widely used open source tools. 

In summary, it can be said  that there is no full fledged power system analysis tool focused on symmetric power systems analysis, which is easy to use and well suited for automation in scientific applications. To fill this gap, we introduce the open source tool \texttt{pandapower} in its current version \textsc{1.4.3}.

\subsection{Introducing pandapower}
\texttt{pandapower} is implemented in Python, guaranteeing free availability and flexible expansion with other open source libraries. Since it is developed as a cross-platform library, it can be deployed seamlessly on computational clusters and parallelized without any license constraints. All implementations are thoroughly verified and wherever possible validated by comparing with commercial software tools. \texttt{pandapower} has been successfully applied in multiple grid studies \cite{Thurner.2017, Scheidler.2017, Menke.2018, india.2017}. Because of the comprehensive model library and the easy to use Python interface, \texttt{pandapower} has a relatively low entry barrier for new users to perform basic power systems analysis, which also makes it a great tool for educational purposes. While \texttt{pandapower} was originally implemented for the analysis of symmetric distribution systems, it has been subsequently extended with models for transmission systems, such as three-winding transformers, shunt elements or network equivalent models. In its current version, \texttt{pandapower} is suited for the analysis of symmetric distribution as well as transmission systems.

\subsection{Network Models} \label{sec:BBM}
Any electric power system analysis function, like power flow or short circuit calculation, is based on a mathematical model of the electric network. There are different approaches how power system tools allow the user to specify this model. A commonly used approach is the bus-branch model (BBM), which defines the network as a collection of buses which are connected by generic branches. Branches are modeled with a predefined equivalent circuit and are used to model multi-pole elements like lines or transformers. Buses are attributed with power injections or shunt admittances to model single-pole elements like loads, generators or capacitor banks. Since the BBM is an accurate mathematical representation of the network, electric equations for power systems analysis can be directly derived from it. The BBM can also be freely parametrized and is not bound to specific models of electric utilities. On the other hand, the user needs to calculate the impedances for each branch and summed power injections at each bus manually from the nameplate data of the grid elements. This can be cumbersome and error-prone especially for complex elements, like transformers with tap changers or more than two windings.

Instead of a BBM, pandapower uses an element-based model (EBM) to model electric grids. An element is either connected to one or multiple buses and is defined with characteristic parameters. Instead of a generic branch model, there are separate models for lines, two-winding, three-winding transformers etc. This allows defining the network with nameplate parameters, such as length and relative impedance for lines, or short circuit voltage and rated apparent power for transformers. Where a BBM allows only the definition of a summed power injection at each bus, single-pole elements (such as load or generation elements) can be connected to buses independently. This also allows connecting multiple elements at one bus. 

The element models need to be processed with the appropriate equivalent circuits to derive a mathematical description of the network. Decoupling the element model from the electric model allows to specify different equivalent circuits for different analysis functionalities. For example, an external grid element can be modeled as slack node in the power flow calculation, but as a voltage source with internal impedance in the short circuit calculation. The EBM also allows composite models that are internally represented by more than one branch, such as a three-winding transformer, or by a combination of bus and branch attributes, such as ward equivalents.

While grid data can be conveniently converted between two tools that use an EBM, converting from a EBM based tool to a BBM requires translation of all elements into their equivalent circuit models. The conversion also leads to a loss of information, as the BBM only includes the electric impedances and no nameplate-information. Since all state-of-the art commercial power systems analysis tools use an EBM, it is much easier to convert grid data from a commercial tool into an open source tool that is also based on an EBM.

\subsection{Data Structure} \label{sec:DataStructure}
\texttt{pandapower} is based on a tabular data structure, where every element type is represented by a table that holds all parameters for a specific element and a result table which contains the element specific results of the different analysis methods. The tabular data structure is based on the Python library pandas \cite{pandas}. It allows storing variables of any data type, so that electrical parameters can be stored together with status variables and meta-data, such as names or descriptions. The tables can be easily expanded and customized by adding new columns without influencing the \texttt{pandapower} functionality. All inherent pandas methods can be used to efficiently read, write and analyze the network and results data.
\begin{figure}[t]
  \centering
	\includegraphics[width=.95\textwidth]{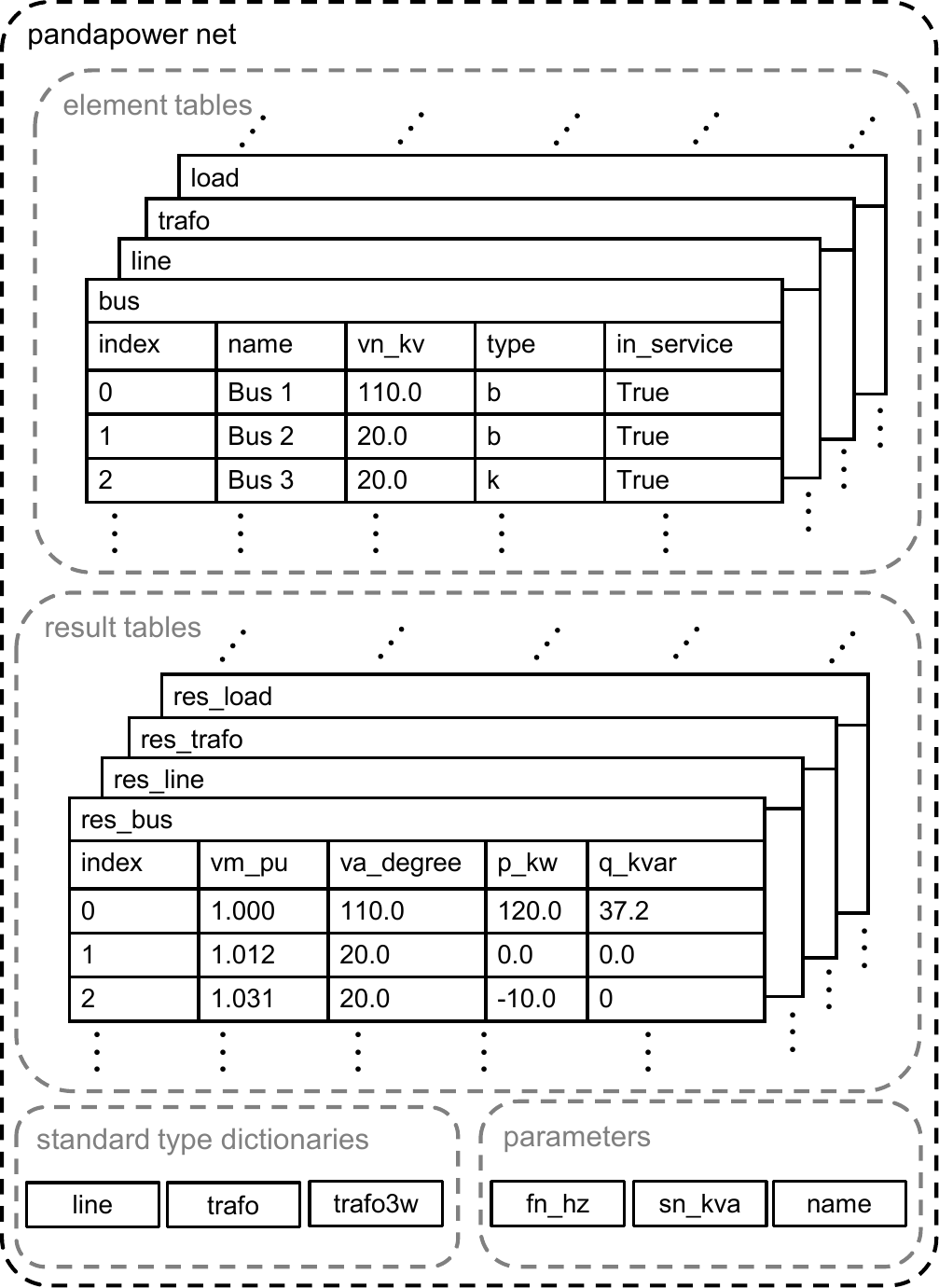}
  \caption{Schematic overview of the \texttt{pandapower} network representation\label{fig:datastructure}}
\end{figure}
A \texttt{pandapower} network (abbreviated as \texttt{net}) is a Python dictionary that holds all information about the network (see Fig.~\ref{fig:datastructure}). Most importantly, it includes an element and a result table for each element type, such as line, transformer, switch etc. (see Section \ref{sec:Elements}). The element table holds all input parameters that are specified by the user, while the result table is used by power flow or optimal power flow functions to store the results. Input and output parameters are identified by the same index in both tables. The \texttt{net} furthermore includes dictionaries which hold standard type data (see Section \ref{sec:stdTypes}) and network wide parameters like frequency, network name or rated apparent power for the per unit system.

\section{Electric Element Models} \label{sec:Elements}
The \texttt{pandapower} library includes many different electric models, some of which are not available in any other open source tool (see Table~\ref{tab:elements}). The electric models and equivalent circuits, representing the different elements, are described in this Section. Detailed formulas for the calculation of the electric parameters are also available in the \texttt{pandapower} documentation \cite{pandapower_rtd}. To allow a convenient step-by-step definition of networks, \textit{create} functions exist for each element.

\subsection{Bus (\texttt{bus})}
Buses represent the nodes of the network. They are defined by a nominal voltage \texttt{bus.vn\_kv}\footnote{All \texttt{pandapower} model parameters are notated as  \texttt{element.parameter} in this paper, where element is the element table in the data structure and parameter the name of the parameter by which it can be accessed. By convention, all parameter names in \texttt{pandapower} end with the parameter unit.}, which is the reference voltage for the per unit system. The rated power for the per unit system is defined system wide with the parameter \texttt{net.sn\_kva}. The voltage magnitude \texttt{res\_bus.vm\_pu} and angle \texttt{res\_bus.va\_degree} are results of a grid analysis.

\subsection{Load (\texttt{load})}
Loads are used to model electric consumption. They are defined by the active power \texttt{load.p\_kw} and reactive power \texttt{load.q\_kvar}. The ZIP model allows modeling loads with constant power, constant current or constant impedance. The percentage of the load which consumes a constant current is defined by the parameter \texttt{load.const\_i\_percent}, the constant impedance part is defined by the parameter \texttt{load.const\_z\_percent}. The rest of the load is assumed to be a constant power load. For constant current and constant impedance, the active power value is assumed to be the power consumption at rated voltage. All nodal powers are noted in passive sign convention (PSC). The load model includes a scaling factor \texttt{load.scaling} that allows to scale the load. 

\subsection{Static Generator (\texttt{sgen})}
Static generators are used to model constant power injection with active power \texttt{sgen.p\_kw} and reactive power \texttt{sgen.q\_kvar}. Since all nodal power is noted in PSC, the power generation is defined to be negative. This might seem unintuitive for generator type elements, but the consistent convention makes the definition of power values unambiguous even for elements where the signing is not obvious, such as external grids, shunts or buses.
The static generator model includes a scaling factor \texttt{sgen.scaling} equivalent to the load scaling factor.

\subsection{Voltage Controlled Generator (\texttt{gen})} \label{sec:gen}
Generator elements are used to model voltage controlled power generation units with a fixed active power injection  \texttt{gen.p\_kw} and a voltage magnitude set point \texttt{gen.vm\_pu}. Adherence with the voltage magnitude in the power flow calculation is achieved by setting the generator bus as a PV node (see Section \ref{sec:PowerFlow}). The reactive power \texttt{res\_gen.q\_kvar} is then calculated so that the voltage magnitude is equal to the set point. Reactive power limits \texttt{gen.q\_min\_kvar} and \texttt{gen.q\_max\_kvar} can be enforced in the power flow, in which case the voltage set point might not always be reached.

\subsection{External Grid (\texttt{ext\_grid})}
The external grid element model represents a voltage source with a voltage magnitude \texttt{ext\_grid.vm\_pu} and the corresponding voltage angle \texttt{ext\_grid.va\_degree}. Adherence with the complex voltage set point in the power flow calculation is achieved by setting the generator bus as a slack node (see Section \ref{sec:PowerFlow}). \texttt{pandapower} supports the connection of multiple external grids in galvanically connected network areas.

\subsection{Shunt (\texttt{shunt})} \label{sec:shunt}
Shunts are network elements that can be used to model a capacitor bank or a reactor. Shunts are defined by a reactive power \texttt{shunt.q\_kvar} and an active power \texttt{shunt.p\_kw} value, which represents the losses. The power values equal the consumption at rated voltage \texttt{shunt.vn\_kv}. The parameter \texttt{shunt.step} allows to model a discretely segmented shunt, such as a switchable capacitor bank.
\begin{table}[!t]
\caption{Comparison of Open Source Element Model Libraries}
\label{tab:elements}
\centering
\begin{tabular}{lrrrrrrr|r}
 & \begin{sideways}MATPOWER 6.0\end{sideways} & \begin{sideways}PYPOWER 5.1.2\end{sideways}  & \begin{sideways}PSAT 2.1.10\end{sideways} & \begin{sideways}OpenDSS 7.6.5\end{sideways} & \begin{sideways}PyPSA 0.10\end{sideways} & \begin{sideways}GridCal\end{sideways} & \begin{sideways}GridLAB-D 3.2\end{sideways} & \begin{sideways}\textbf{pandapower 1.4.3}\end{sideways} \bigstrut\\
\hline
ZIP-load &  &  & \checkmark    & \checkmark   &      & \checkmark &    \checkmark & \checkmark \bigstrut\\ \hline
Line & \checkmark & \checkmark & \checkmark     & \checkmark     & \checkmark     & \checkmark      & \checkmark & \checkmark \bigstrut\\ \hline
2-Winding Transformer ($\pi$) & \checkmark & \checkmark    & \checkmark     & \checkmark     & \checkmark     & \checkmark      & \checkmark & \checkmark \bigstrut\\ \hline
2-Winding Transformer (T) & &     &      &  \checkmark   & \checkmark     &    & \checkmark & \checkmark \bigstrut\\ \hline
3-Winding Transformer &  &  & \checkmark     &  \checkmark     &       &          &  \checkmark & \checkmark \bigstrut\\ \hline
DC Line & \checkmark &   & \checkmark     & \checkmark     & \checkmark      &    &   \checkmark   & \checkmark \bigstrut\\ \hline
Ideal Switches &  &     &    &       &       &          &   & \checkmark \bigstrut\\ \hline
Volt.-Controlled Generator & \checkmark & \checkmark    & \checkmark     & \checkmark     & \checkmark  & \checkmark      & \checkmark  & \checkmark \bigstrut\\ \hline
Static Load / Generation & \checkmark & \checkmark     & \checkmark     & \checkmark     &   \checkmark    & \checkmark      &     \checkmark & \checkmark \bigstrut\\ \hline
Shunt & \checkmark  & \checkmark   & \checkmark     & \checkmark     & \checkmark     & \checkmark        &  \checkmark & \checkmark \bigstrut\\ \hline
Asymmetrical Impedance &  &  &      &      &       &         &   & \checkmark \bigstrut\\ \hline
Ward Equivalents &    &   &       &       &       &         &    & \checkmark  \bigstrut\\ \hline
Storage Unit &   &    &     &   \checkmark    & \checkmark      &            & \checkmark &  \bigstrut\\
\end{tabular}
\end{table}
\subsection{Line (\texttt{line})} \label{sec:line}
Lines are modeled with a $\pi$-equivalent circuit~\cite{BrownBook}. The electric parameters of a line are specified relative to the length of the line \texttt{line.length\_km}. The longitudinal impedance is defined by the resistance \texttt{line.r\_ohm\_per\_km} and reactance \texttt{line.x\_ohm\_per\_km}. The shunt admittance is defined by the capacity \texttt{line.c\_nf\_per\_km}.
The line  current \texttt{res\_line.i\_ka} is calculated as the maximum current at both ends of the line. The line loading \texttt{res\_line.loading\_percent} can be calculated as a ratio of line current \texttt{res\_line.i\_ka} to the maximum thermal line current \texttt{line.max\_i\_ka}. A derating factor \texttt{line.df} can be defined to consider the fact that some lines might not be utilized to their full thermal capacity. The model also provides a parameter \texttt{line.parallel} to define the number of parallel lines.
\begin{table*}[!t]
\renewcommand{\arraystretch}{1.0}
\caption{Switch configurations with common approximation and internal representation in \texttt{pandapower} switch model \label{fig:switches}}
\begin{tabular}{p{0.14\linewidth}|p{0.2\linewidth}|p{0.2\linewidth}|p{0.2\linewidth}}
& Bus-Bus Switches & Bus-Line Switches & Bus-Transformer Switches \\ \hline
\mbox{Switch} \mbox{Configuration} &
\multirow{3}{*}{\includegraphics[width=1.0\linewidth]{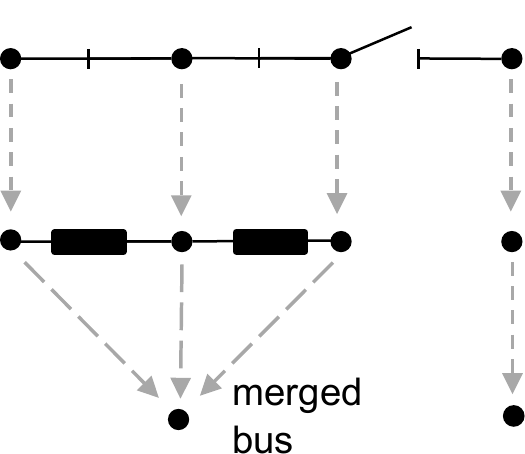}} & 
 \multirow{3}{*}{\includegraphics[width=1.0\linewidth]{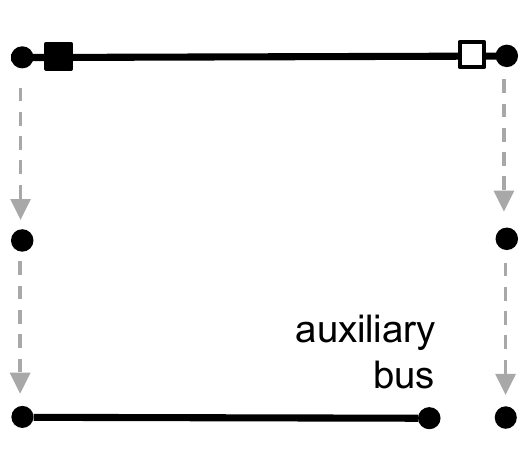}} & 
\multirow{3}{*}{\includegraphics[width=1.0\linewidth]{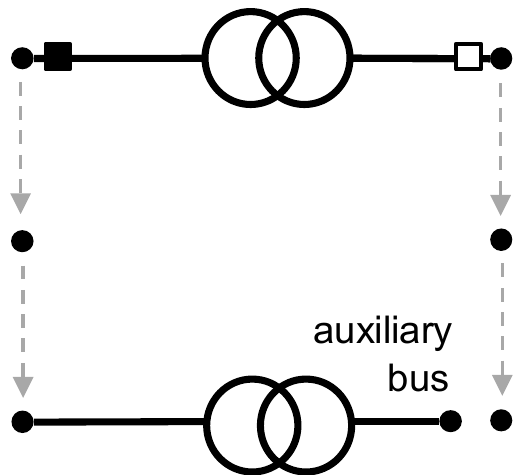}} \\[2.2em] \hline
 \mbox{Common} \mbox{Approximation} & & & \\[2.2em] \hline
 \mbox{pandapower} \mbox{Switch Model} & & & \\[2.2em] 
\end{tabular}
\end{table*}

\subsection{Two-Winding Transformer (\texttt{trafo})} \label{sec:trafo}
Two-winding transformers are commonly modeled with a $T$-equivalent circuit \cite{BrownBook}. However, for the sake of completeness, \texttt{pandapower} also includes a $\pi$-transformer model. The longitudinal impedance is defined by the short circuit voltage \texttt{trafo.v\_sc\_percent} and its real part \texttt{trafo.v\_scr\_percent}. The real part of the transformer impedance represents the copper losses in the transformer windings. The shunt admittance represents the losses in the iron core of the transformer. The open loop current \texttt{trafo.i0\_percent} defines the overall open loop losses and an active power loss \texttt{trafo.pfe\_kw} defines the iron losses. The rated transformer voltages for the high voltage side \texttt{trafo.vn\_hv\_kv} and the low voltage side \texttt{trafo.vn\_lv\_kv} define the nominal transformer ratio and do not necessarily have to be equal to the rated voltages of the connected buses. If an angle shift \texttt{trafo.shift\_percent} is defined, the ratio becomes complex and the voltage angle between high and low voltage side is shifted. The transformer ratio can also be influenced by defining a tap changer and its current position \texttt{trafo.tp\_pos}. With every step the tap position \texttt{trafo.tp\_pos} diverges from its medium position \texttt{trafo.tp\_mid}, the transformer ratio changes by a percentage defined by \texttt{trafo.tp\_st\_percent}. It is also possible to define an angle shift per step \texttt{trafo.tp\_degree\_percent} to model phase shifting transformers. The tap changer can be  located at the low voltage or the high voltage side of the transformer which is defined by the parameter \texttt{trafo.tp\_side}.
The loading \texttt{res\_trafo.loading\_percent} is calculated of the maximum loading at high and low voltage side. It can either be calculated in reference to the nominal power \texttt{trafo.sn\_kva} or to the nominal current.
Just as for lines, there is a parameter \texttt{trafo.parallel} which allows the definition of multiple parallel transformers in one element.

\subsection{Three-Winding Transformer (\texttt{trafo3w})} \label{sec:trafo3w}
Three-winding transformers can be modeled by three two-winding transformers in wye connection \cite{BrownBook}. The three-winding transformer model in \texttt{pandapower} carries out this conversion internally. The open loop losses defined by \texttt{trafo3w.i0\_percent} and \texttt{trafo3w.pfe\_kw} are considered in the high voltage side transformer. The short circuit voltages of the two-winding transformers are calculated with a wye-delta conversion from the short circuit voltages \texttt{trafo3w.vsc\_hv\_percent}, \texttt{trafo3w.vsc\_mv\_percent} and \texttt{trafo3w.vsc\_lv\_percent} as well as their respective real parts \texttt{trafo3w.vscr\_hv\_percent}, \texttt{trafo3w.vscr\_mv\_percent} and  \texttt{trafo3w.vscr\_lv\_percent}. The equivalent circuit impedances for the three two-winding transformers are then calculated from the nameplate parameters according to the two-winding transformer model. The loading \texttt{res\_trafo3w.loading\_percent} is calculated as the maximum loading of the three two-winding transformers. The loading of the equivalent two-winding transformers is calculated either relative to the rated apparent powers \texttt{trafo3w.sn\_lv\_kva}, \texttt{trafo3w.sn\_mv\_kva} and \texttt{trafo3w.sn\_hv\_kva} or relative to the respective rated current as described in Section \ref{sec:trafo}.

\subsection{Switch (\texttt{switch})}
The switch element allows modeling of ideal switches. A switch element connects a bus  \texttt{switch.bus} with an element \texttt{switch.element}. The element type is defined by the parameter \texttt{switch.et} and can either be a second bus,  a line or a transformer.  The \texttt{switch.closed}  parameter signals if the switch is open or closed.  A closed bus-bus switch galvanically connects two buses without a voltage drop. In network calculation tools without an explicit switch model, bus-bus switches can only be modeled as a small impedance between two buses (see Table~\ref{fig:switches}). This can however lead to unwanted voltage drops and convergence problems in the power flow. The \texttt{pandapower} switch model avoids this problem by internally fusing buses that are connected by closed bus-bus switches as shown in Table~\ref{fig:switches}. Branches that are connected to a bus through an open switch are often modeled by neglecting or disabling the branch element (see Table~\ref{fig:switches}). This however means that the information about the switch position is lost and the open loop current of the branch element is neglected. \texttt{pandapower} instead internally switches the branch over to an auxiliary bus so that the branch is disconnected from the bus but the loading current is still considered. 

\subsection{DC Transmission Line (\texttt{dcline})}
A DC transmission line transmits active power between two buses. The transmitted active power \texttt{dcline.p\_kw} is reduced by absolute transformation losses \texttt{dcline.p\_loss\_kw} and relative transmission losses \texttt{dcline.p\_loss\_percent} at the destination bus. A DC line is modeled with two generators at both buses, where the voltage control with reactive power works just as described for the generator model in Section \ref{sec:gen}.

\subsection{Impedance (\texttt{impedance})}
An impedance element connects two buses with a per unit impedance in relation to the rated power \texttt{impedance.sn\_kva}. The impedance does not have to be symmetrical, in which case the nodal point admittance matrix becomes asymmetrical. The forward impedance $\underline{z}_{ft}$ is defined by \texttt{impedance.rft\_pu} and \texttt{impedance.xft\_pu}, the backward impedance $\underline{z}_{tf}$ by \texttt{impedance.rft\_pu} and \texttt{impedance.xft\_pu}. Asymmetrical impedances are used as equivalent elements in network reduction.

\subsection{Ward Equivalents (\texttt{ward} / \texttt{xward})}  \label{sec:ward}
The Ward equivalent is a combination of a constant apparent power consumption and a constant impedance load \cite{ward}. The constant impedance load is given as active and reactive power consumption \texttt{ward.pz\_kw} and \texttt{ward.qz\_kvar} at the rated voltage of the bus. The constant active and reactive power is given by \texttt{ward.ps\_kw} and \texttt{ward.qs\_kvar}. The extended Ward equivalent includes an additional voltage source with internal impedance \cite{ward}. The voltage source is modeled as a generator with zero active power and a voltage set point defined by the parameter \texttt{xward.vm\_pu}. The internal impedance is defined by the parameters \texttt{xward.r\_ohm} and \texttt{xward.x\_ohm}.

\section{Electric Network Analysis} \label{sec:electric_analysis}
\begin{figure}[tp]
\centering
\includegraphics[width=\linewidth]{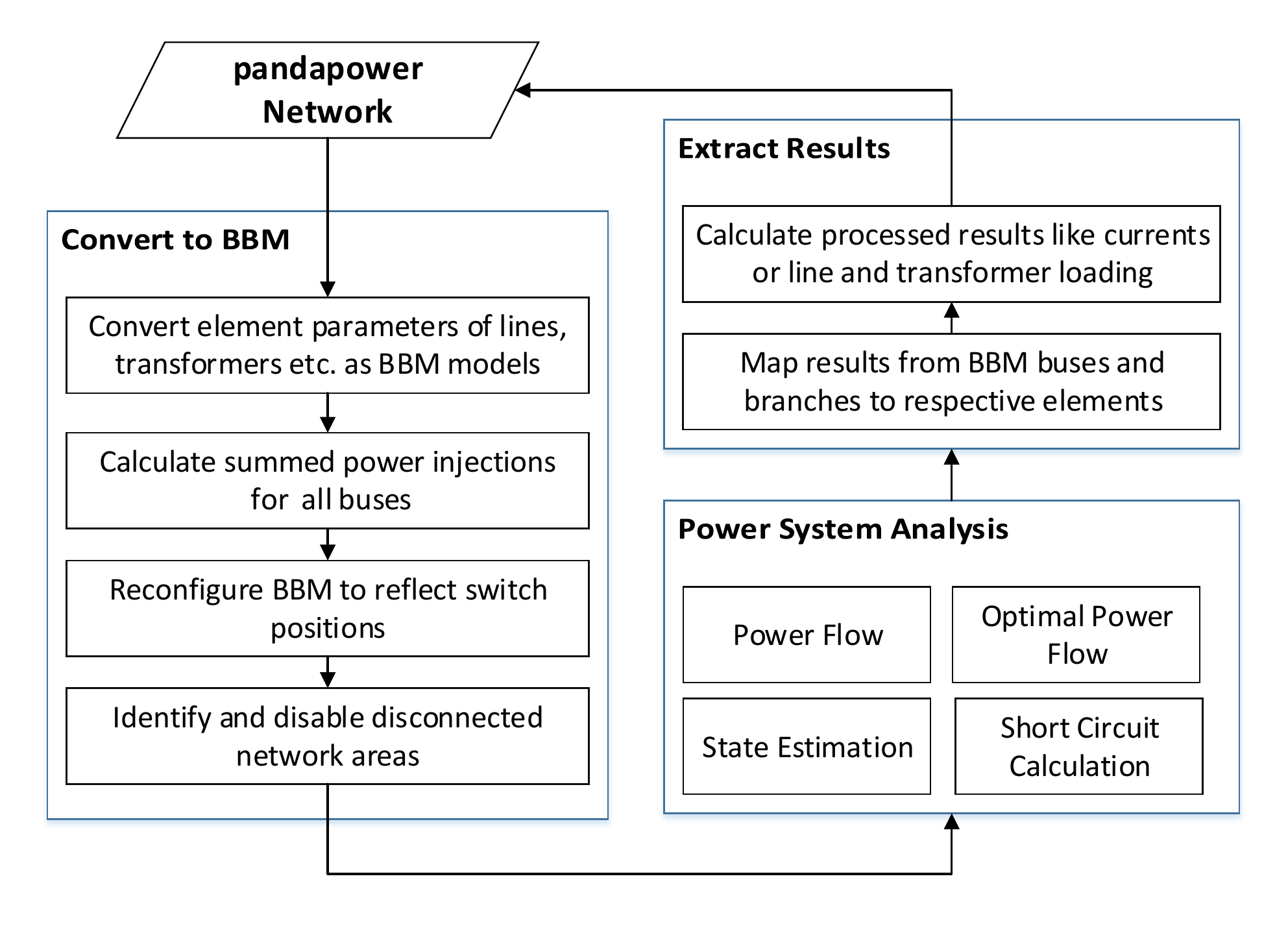}
\caption{Electric power system analysis in \texttt{pandapower} \label{fig:electric_analysis}}
\end{figure}

With the possibility to conduct power flow, optimal power flow, state estimation and short circuit calculations, \texttt{pandapower} provides all of the most commonly used static network analysis features. As outlined in Section \ref{sec:DataStructure}, the \texttt{pandapower} data structure contains common nameplate parameters for convenient parametrization. To carry out electric network analysis, all element models have to be translated into their electric equivalent representation. This translation is done by converting the element-based data structure internally into a BBM as shown in Fig.~\ref{fig:electric_analysis}. The internal BBM model has a similar structure as a \textsc{pypower} casefile, but has been extended to include parameters like asymmetrical impedances or constant current load. The correlation between \texttt{pandapower} elements and the BBM is not trivial for multiple reasons. First, the indexes for the BBM matrices have to be continuous starting from zero, while the indexes in the pandas tables can be unsorted and intermittent. Second, some \texttt{pandapower} elements translate into multiple buses and branches (e.g. three-winding transformers, extended ward equivalents). Third, multiple elements (e.g. loads, shunts, generators or ward elements) can be connected to the same bus in \texttt{pandapower}, so that properties in the BBM for a bus can possibly originate from different \texttt{pandapower} element tables. Fourth, to dissolve switches as shown in Table~\ref{fig:switches}, it is necessary to create new auxiliary buses, reconnect branches and merge multiple \texttt{pandapower} buses into one BBM bus. And fifth, areas which have no galvanic connection to any slack bus are identified and disabled in the BBM, so that some elements might be enabled in \texttt{pandapower} but disabled in the BBM. To keep track of the complex relationship between the \texttt{pandapower} elements and their representations in the BBM, several mappings are created during the conversion process. After the electric analysis is conducted based on the BBM, the obtained results are allocated to the elements accordingt to these mappings. In this way, the results can also be set in relation to the input parameters, for example to calculate line loadings as a ratio of the maximal thermal current and the actual current resulting from the power flow. 

\subsection{Power Flow} \label{sec:PowerFlow}
The \texttt{pandapower} power flow solver is based on the Newton-Raphson method \cite{grainger1994power}. The implementation was originally based on \textsc{pypower}, but has been improved with respect to robustness, runtime and usability.

\begin{figure}[t]
  \centering
  \includegraphics[width=.85\textwidth]{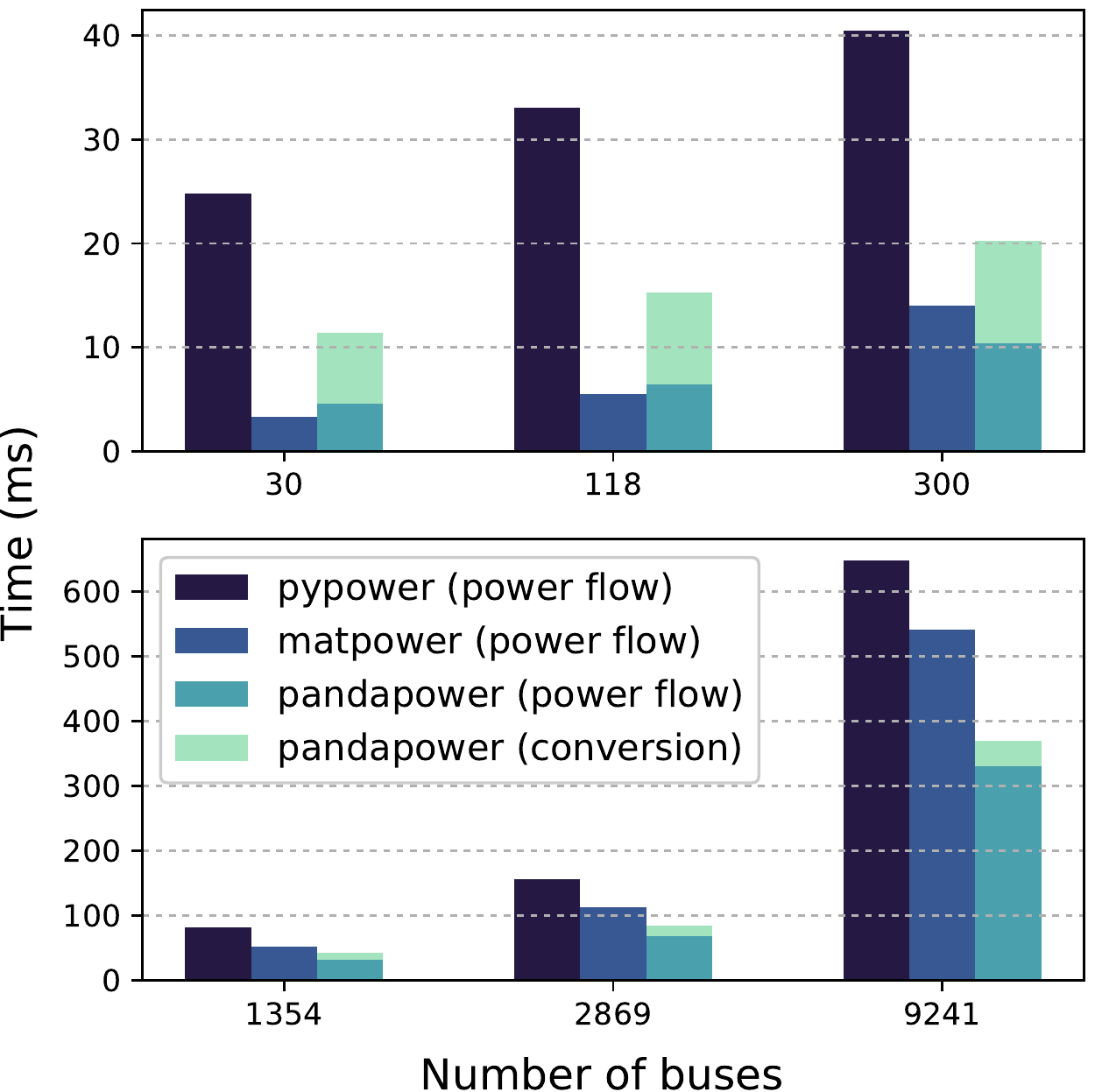}
  \caption{Speed comparison of \texttt{pandapower}, \textsc{pypower} and \textsc{matpower} for \textsc{matpower} casefiles \label{fig:speed_comparison}}
\end{figure}

Internal power flow parameters, such as node type for the power flow calculation (slack, PV or PQ node) or per unit conversions, are carried out automatically by \texttt{pandapower}. This improves user convenience and reduces the risk of incoherent input data. \texttt{pandapower} offers three different methods to initialize the complex voltage vector for the AC power flow calculation. It can either be the result of a previous power flow calculation, the solution of a DC power flow or a flat start. Initializing with a DC power flow is recommended in meshed networks, where large voltage angle differences between the buses might lead to non-convergence in case of a flat start. In radial distribution grids on the other hand, the reference voltage angle is dictated by the external grid so that relative voltage angle shifts of transformers have no impact on the power flow result. That is why \texttt{pandapower} provides the option to neglect the voltage angles to allow faster and more robust convergence in radial distribution grids. The additional conversion step that is necessary to convert the \texttt{pandapower} model to a BBM and map back the results afterwards causes an additional overhead compared to programs that operate directly on the BBM, like \textsc{matpower} or \textsc{pypower}. On the other hand, some parts of the \texttt{pandapower} solver have been accelerated using the just-in-time (jit) compiler numba \cite{Lam.2015}. To outline the difference in computational time, Fig.~\ref{fig:speed_comparison} shows the calculation time for different standard \textsc{matpower} case files. The displayed timings are the shortest of 100 loops of a power flow calculation to minimize the influence of other processes running on the benchmark system. A flat start is chosen for all three tools to have the same initial conditions. The \texttt{pandapower} timings distinguish between power flow solver and conversion overhead, which includes BBM conversion as well as result extraction. It can be seen that \texttt{pandapower} is faster than \textsc{pypower} in all cases due to the jit accelerated building of the Jacobian matrix and other aspects of the Newton-Raphson solver. It can also be seen that while the conversion overhead takes up more than half of the calculation time for small networks, its share decreases significantly for larger networks. While \texttt{pandapower} is slower than \textsc{matpower} for small networks, it is faster for medium sized and large networks, even including the conversion overhead for the BBM. By default, the BBM conversion is carried out before every power flow. However, if multiple subsequent power flows are performed for the same network that only differ in the nodal power injections, the conversion into a BBM becomes redundant. For this reason, \texttt{pandapower} offers the possibility to reuse the BBM and the nodal point admittance matrix from previous power flow calculations. This feature can speed up applications like quasi-static time series simulations or heuristic power set point optimizations. In addition to the default Newton-Raphson solver, \texttt{pandapower} also provides an implementation of a backward/forward sweep \cite{sweep}. It is also possible to use the fast decoupled as well as the Gauss-Seidel power flow algorithms through an interface to \textsc{pypower}.

\subsection{Optimal Power Flow} \label{sec:OPF}
\texttt{pandapower} allows solving AC and DC optimal power flow (OPF) problems through interfacing \textsc{pypower}. The interior point solver \cite{wang2007computational, wang2007computation} provided by \textsc{pypower} is used to solve the problem, while costs, flexibilities and constraints are configured through the element-based \texttt{pandapower} data structure. This allows all electric element models provided by \texttt{pandapower} to be used in the OPF. Branch constraints are given as maximum loading for transformers and lines, instead of absolute limits for power flows. Bus constraints include maximum and minimum voltage magnitude. Active and reactive power limits can be defined for PV/slack-elements like external grids and generators, but also for PQ-elements, such as loads and static generators. This allows flexible consideration of static generators in dispatch optimizations as well as the consideration of load shedding. The cost function for each power injection or load can either be defined by a piecewise linear or a n-polynomial cost function of the active and reactive power output of the respective elements.

\subsection{State Estimation}\label{sec:StateEst}
\texttt{pandapower} includes a state estimation module that allows to estimate the electrical state of a network by dealing with inaccuracies and errors from measurement data. The weighted-least-squares optimization algorithm minimizes the weighted squared differences between measured values and the corresponding power flow equations \cite{abur2004power}.

\texttt{pandapower} supports bus, line and transformer measurements. Bus measurements can be given for voltage magnitude or active and reactive power injections. Measurements at lines or transformers can be given for current magnitude or active and reactive power flows at either end of the branch.

The state estimation may not converge if measurements include bad data. Therefore, it is necessary to remove bad data prior to the estimation process. This problem is solved in \texttt{pandapower} with a $\chi^2$ test and a normalized residual test \cite{abur2004power}. A $\chi^2$ test is able to identify the probability that bad measurements exist in the measurement set or if the network topology does not fit the measurement data. A normalized residuals test can take information of the $\chi^2$ test, compute the normalized residuals and remove the measurement with the highest residual. The cycle is repeated until the bad data check passes or no measurements can be removed any more.

\subsection{Short Circuit Calculation according to IEC 60909}\label{sec:SC}
While short circuit currents are an inherently transient phenomenon, they can be approximated based on a static network model. The IEC~60909 standard \cite{iec60909} defines rules to calculate certain characteristic values of the short circuit, such as the initial short circuit current $I_k^{''}$, peak short circuit current $i_p$ or long term SC current $I_k$. The calculation of initial sub-transient short circuit currents for symmetrical three-phase short circuits as well as two-phase short circuits  is implemented in \texttt{pandapower}. The necessary correction factors are implemented in \texttt{pandapower} according to the standard and are automatically applied in the conversion to the BBM. Additional input parameters, which are necessary to calculate internal impedances of external grids or synchronous generators, are defined in the element tables, together with the default parameters. The implementation allows modeling power converter elements, such as PV plants or wind parks, as constant current sources according to the 2016 revision of the standard \cite{iec60909}. 

\section{Topological Network Analysis}\label{sec:Top}
\texttt{pandapower} provides the possibility of graph searches using the Python library NetworkX \cite{networkx} by providing a possibility to translate \texttt{pandapower} networks into NetworkX graphs. Once a network is translated into an abstract graph, all graph searches implemented in the NetworkX library can be used to analyze the network structure. It is then possible for example to find connected components or cycles in the graph and transfer the results back to \texttt{pandapower}. The line length can be translated as edge weight in the graph so that it is possible to find the shortest path between two buses or measure distances between buses in the network. The translation of the network into a graph can also be configured depending on the use case. For example, lines with open switches are not transferred as edges into the graph by default, since there is no electric connection between those nodes. If a graph search is however aimed at the physical, rather than the electrical, structure, it might be desired to include those branches into the translation as well. Additionally, \texttt{pandapower} also provides some predefined search algorithms to tackle common graph search problems in electric networks, such as finding all unsupplied buses, finding galvanically connected buses or identifying buses on main or secondary network feeders.

\begin{figure}[t]
  \centering
  \includegraphics[width=.85\textwidth]{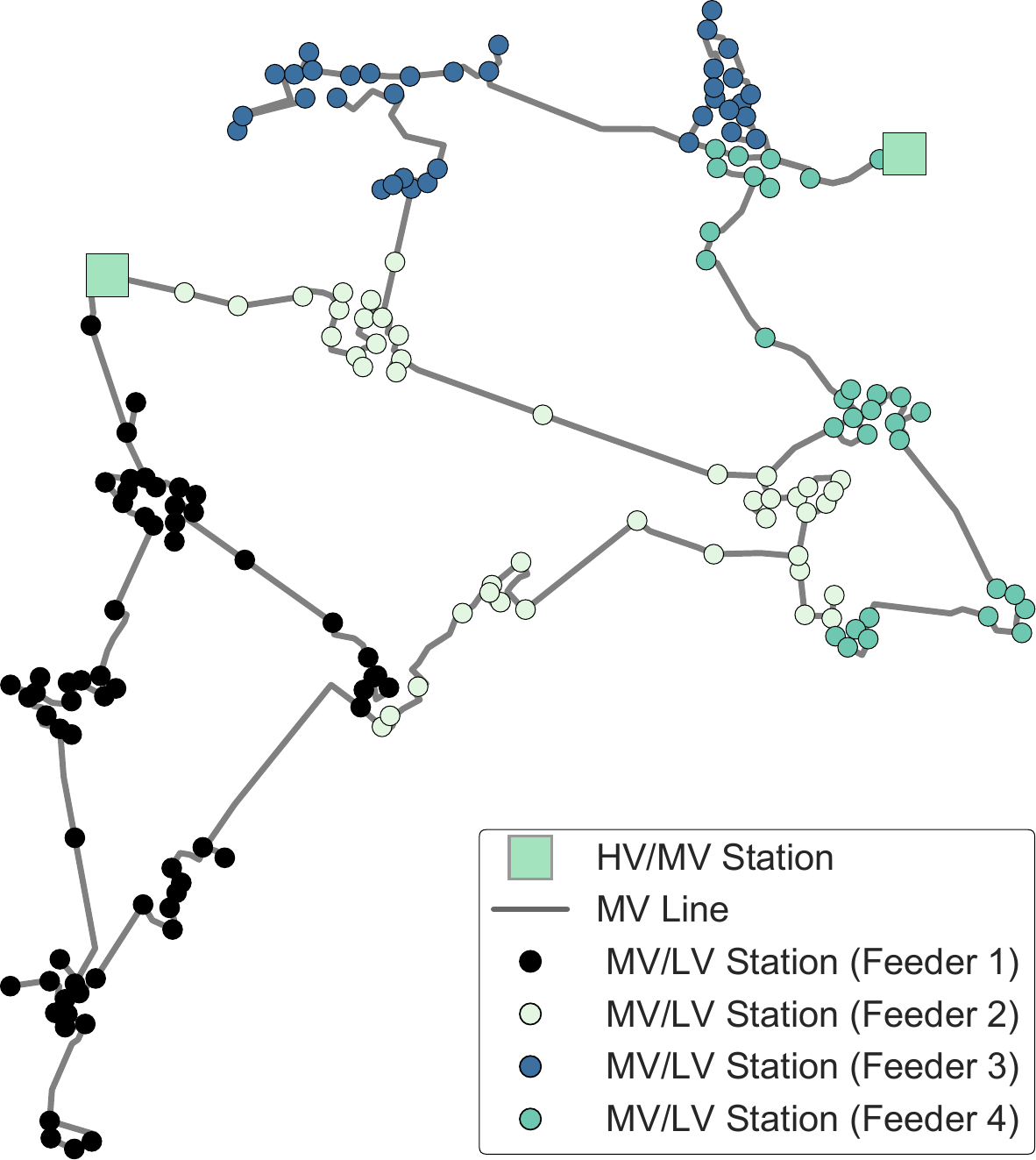}
  \caption{Plot with stations colored per feeder to highlight the radiality of the generic \texttt{pandapower} MV Oberrhein network \label{fig:plot}}
\end{figure}

\section{Further Functionality}

\subsubsection{Standard Type Libraries} \label{sec:stdTypes}
Lines and transformers have two different categories of parameters: parameters that depend on the specific element (e.g. the length of a line or the bus to which a transformer is connected to) and parameters that only depend on the type of line or transformer which is used (e.g. the rated power of a transformer or the resistance per kilometer line). \texttt{pandapower} includes a standard type library that allows the creation of lines and transformers using predefined basic standard type parameters. The user can either define individual standard types or use the predefined \texttt{pandapower} basic standard types for convenient definition of networks.

\subsubsection{Predefined Networks}
In addition to creating networks through the application programming interface (API), 66 predefined, published test and benchmark networks can be directly accessed through \texttt{pandapower}. These include the well-known IEEE power system test cases \cite{matpower}, benchmark networks from CIGRE \cite{cigre} as well as generic medium and low voltage networks.

\subsubsection{Plotting}
\texttt{pandapower} comes with extensive plotting features using the matplotlib library \cite{Hunter:2007}. All \texttt{pandapower} elements can be translated into different matplotlib collections that can be customized with respect to shape, size and color to allow highlighting and create individual network plots. It is also possible to use colormaps to codify information, like the loading of lines or the voltage at buses. An example plot of a generic \texttt{pandapower} MV network is shown in Fig.~\ref{fig:plot}.

In addition, networks plots through plotly \cite{plotly} are also supported, which allows interactive features, such as element selection or displaying hovering information.

\subsubsection{Converter} \texttt{pandapower} includes converters in order to export a \texttt{pandapower} grid as a \textsc{matpower} or \textsc{pypower} case file or to import a casefile to a \texttt{pandapower} grid. When importing, the converter tries to guess the parameters of the grid as best as possible, but there is always a loss of information since \textsc{matpower} and \textsc{pypower} are not element based. For example, if only the impedance is known, it is simply not possible to deduce both the relative impedance as well as the line length.

\section{Case Study} \label{sec:case_study}
An exemplary case study is carried out to showcase the capabilities of \texttt{pandapower} with respect to grid modeling and analysis. It shows how the different network analysis functionalities of \texttt{pandapower} can be easily combined with minimal effort to conduct an investigation that can hardly be accomplished with commercial software and would require a lot more code in other existing open source frameworks. The case study simulates an active grid operation in a radial grid topology with a quasi-static time series simulation. The implementation is available as an interactive tutorial, via Jupyter Notebook, on github \cite{casestudy}.

\begin{figure}[h]
  \centering
  \includegraphics[width=1.\textwidth]{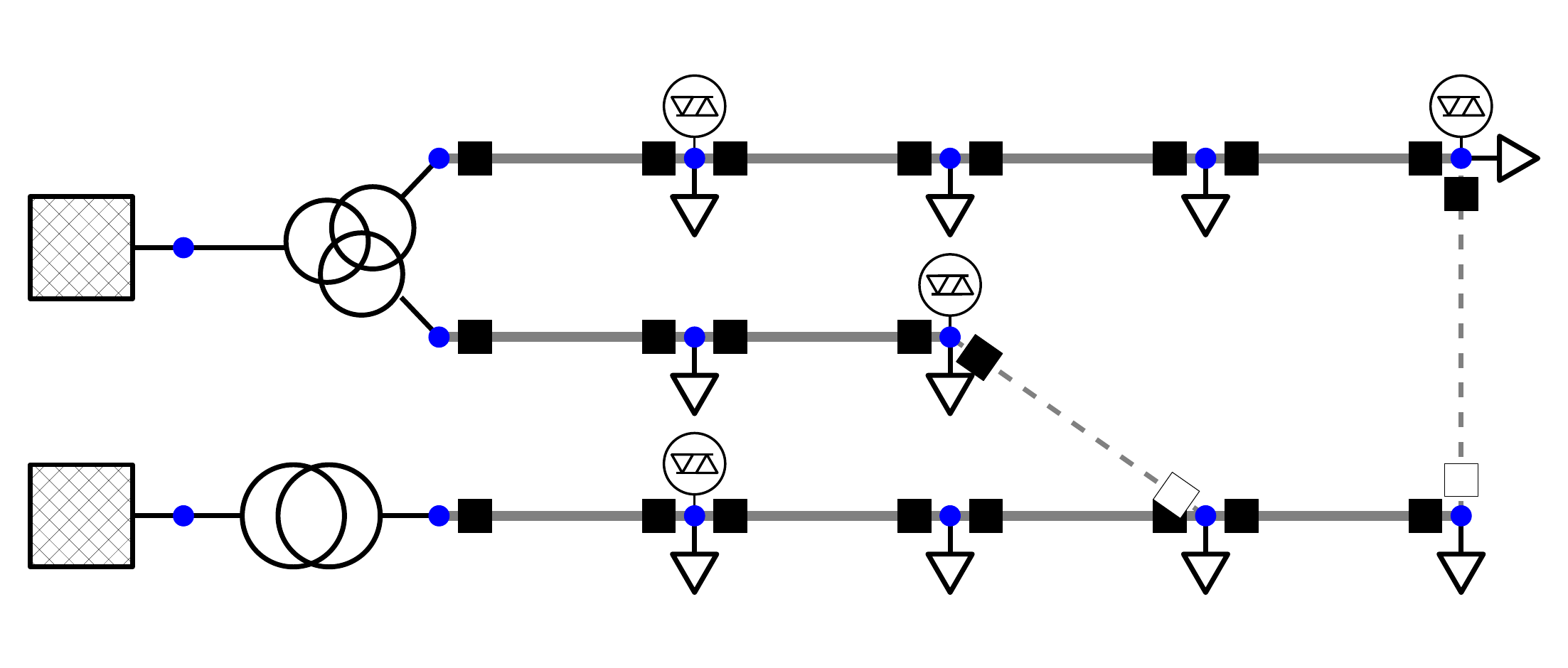}
  \caption{Case study grid \label{fig:case_study:grid}}
\end{figure}

The case study grid, shown in Figure~\ref{fig:case_study:grid}, is a \unit[10]{kV} ring-main grid, fed from two \unit[110]{kV} connections points through a two-winding and a three-winding transformer respectively. 10 loads and 4 wind power systems are connected through 12 lines in a branched feeder layout. Such a grid can be defined in the \texttt{pandapower} format, including all relevant electric information, with only 32 lines of programming code. The grid plot shown in Figure \ref{fig:case_study:grid} is generated with the \texttt{pandapower} plotting module.

\subsection{Grid Operation restrictions} \label{sec:case_study:operation}
The goal of the case study is to simulate active grid operation, including continuous reconfiguration of the switching state as well as the adaption of the transformer tap changer positions to ensure safe operation of the grid. The following exemplary restrictions are formulated with regard to grid operation:
\begin{itemize}
\item The grid has to be operated radially.
\item All bus voltages have to be within $\pm \unit[3]{\%}$ of the rated voltage at all times.
\item All line and transformer loadings have to be below \unit[50]{\%} at all times to allow continued safe operation in case of a contingency.
\item All possible fault currents have to be above \unit[1.1]{kA} at all times, so that they can be clearly distinguished from normal operation currents by the protection system.
\end{itemize}
The goal is to operate the grid with minimal active power losses, while complying with all of these constraints. 

\subsection{Analysis of Switching States}
Since the grid is operated radially, tie-line switches have to be opened to separate the different feeders at all times. Figure~\ref{fig:case_study:grid} shows one permissible switching state. Theoretically, many more possible configurations are possible by the total of 24 tie line switches in the grid. Considering that two switches need to be opened to separate the three feeders, there are a total of $n=\frac{24 \cdot 23}{2}=276$ possible switching states. Not all of theses switching states are permissible, however: switching states, that lead to meshing in the grid or cause stations to be separated from power supply, are not feasible. In this case study, each switching state is analyzed with the \texttt{pandapower} topology package to check its validity. Of the 276 switching states, 96 lead to an infeasible topology, so that 180 potential configurations remain. In addition to the radiality constraint, the switching state also impacts short-circuit currents, which are primarily defined by the length of the feeders. All 180 topologically valid switching states are therefore checked for their minimal fault current. The analysis yields that the minimal fault currents falls below the defined threshold of \unit[1.1]{kA} in 100 cases. That leaves 80 switching states that comply with topological as well as short-circuit constraints, and are therefore considered as potential switching states in the time-series simulation.
A function that checks the validity of each switching state is implemented based on \texttt{pandapower}. It iterates over all switching states and checks topological as well as short-circuit restrictions. The efficient use of the interface to the NetworkX library in the \texttt{pandapower} topology module and the usage of the \texttt{pandapower} shortcircuit module allows to implement this functionality with only 23 lines of code.

\begin{figure}[t]
  \includegraphics[width=1.\textwidth]{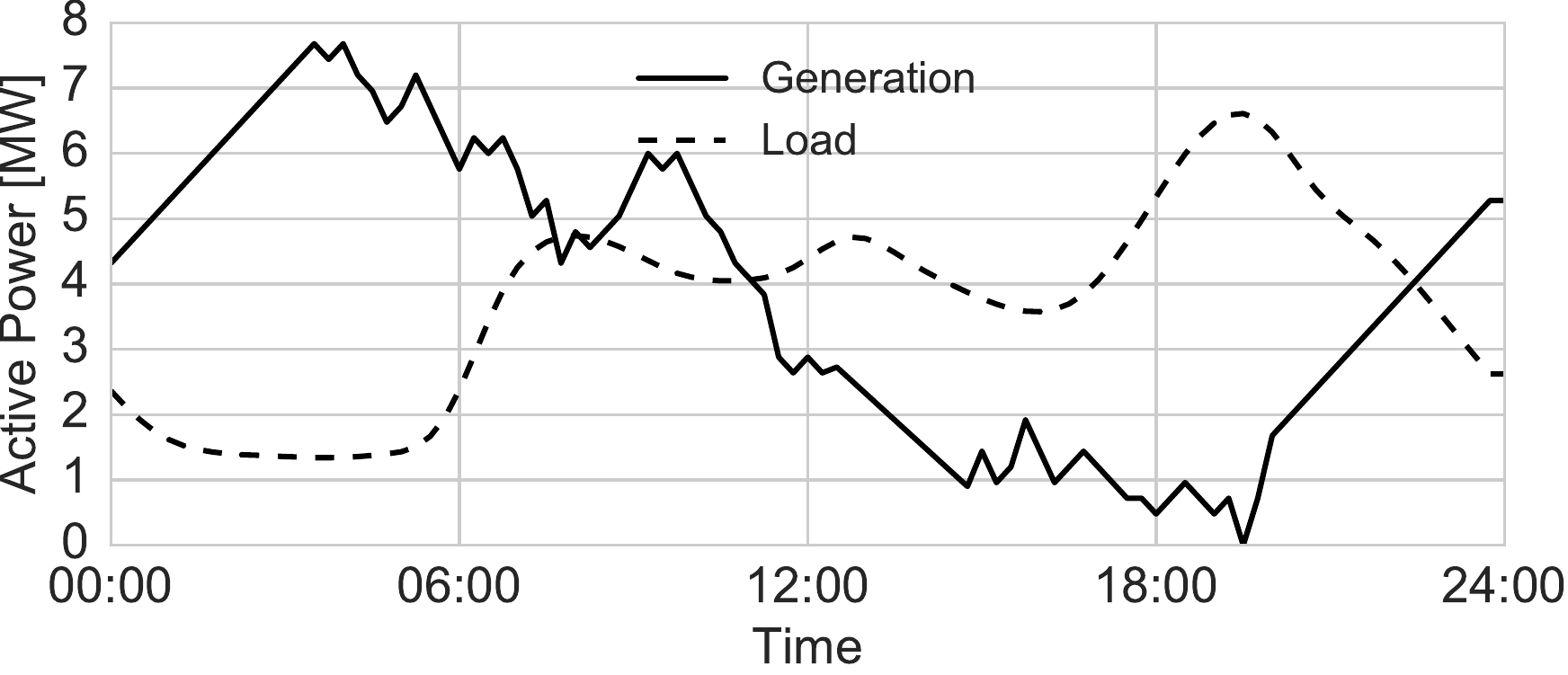}
  \caption{Load and generation profile for an exemplary day \label{fig:case_study:series}}
\end{figure}

\subsection{Time-series simulation}
A quasi-static time-series simulation is carried out with the load and generation profile shown in Figure~\ref{fig:case_study:series}. For each time-step, tap-changer positions as well as the switching state are optimized. The goal is to minimize active power losses, while complying with all constraints defined in Section \ref{sec:case_study:operation}. Radial operation and short-circuit current constraints are already fulfilled for the 80 switching states that have been found to be feasible. This leaves the constraints for bus voltages and line loadings, which depend on the load and generation profile. In each time step, a power flow is therefore carried out for each of the 88 valid switching states, to check adherence to the defined constraints. If any line or transformer surpasses the maximum loading of \unit[50]{\%}, the switching state is considered not to be valid in this specific time step. If the bus voltage constraint is violated, the tap-changer positions are adapted to keep the voltages within the limits. The transformer, which is responsible for the voltage control at a specific bus, depends on the feeder configuration. This transformer is identified with a graph search using the \texttt{pandapower} topology package, if a power flow calculation yields a violation of voltage constraints. The tap position at that transformer is then iteratively adapted until the voltage problem is solved. If the voltage problem cannot be solved through the tap changers, the switching state is also not considered to be valid. Out of all switching states that are found to be valid, the one with the lowest active power losses is chosen as the optimal configuration in this time step.
A function that analyses all switching states for one time-step with regard to power flow constraints, adapts the tap changers so that voltage constraints are complied with, checks for line and transformer overloading and returns the switching state and tap positions that lead to minimal active power losses while complying with all constraints is implemented in pandapower. Based on the \texttt{pandapower} power flow as well as the \texttt{pandapower} topology package, this functionality can be implemented with only 41 lines of code. The function that iterates over the time series and collects the results in each time step is implemented with 7 lines of code. 
\begin{figure}[t]
  \centering
  \includegraphics[width=1.\textwidth]{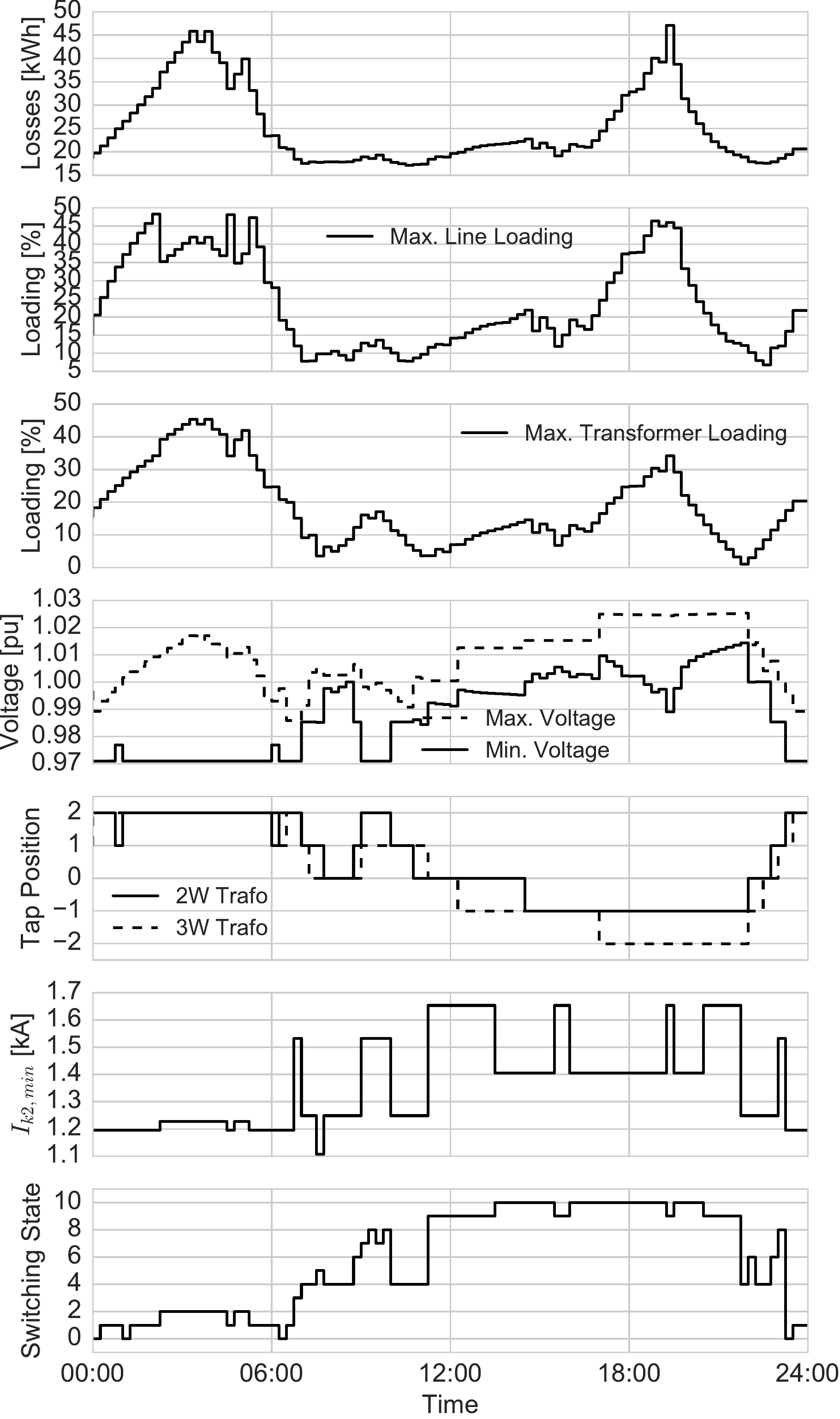}
  \caption{Results of time series simulation \label{fig:case_study:result}}

\end{figure}
\subsection{Results}
\begin{figure}[t]
  \centering
	\subfloat[Grid state for peak generation at 03:30 h]{\includegraphics[width=1.\textwidth]{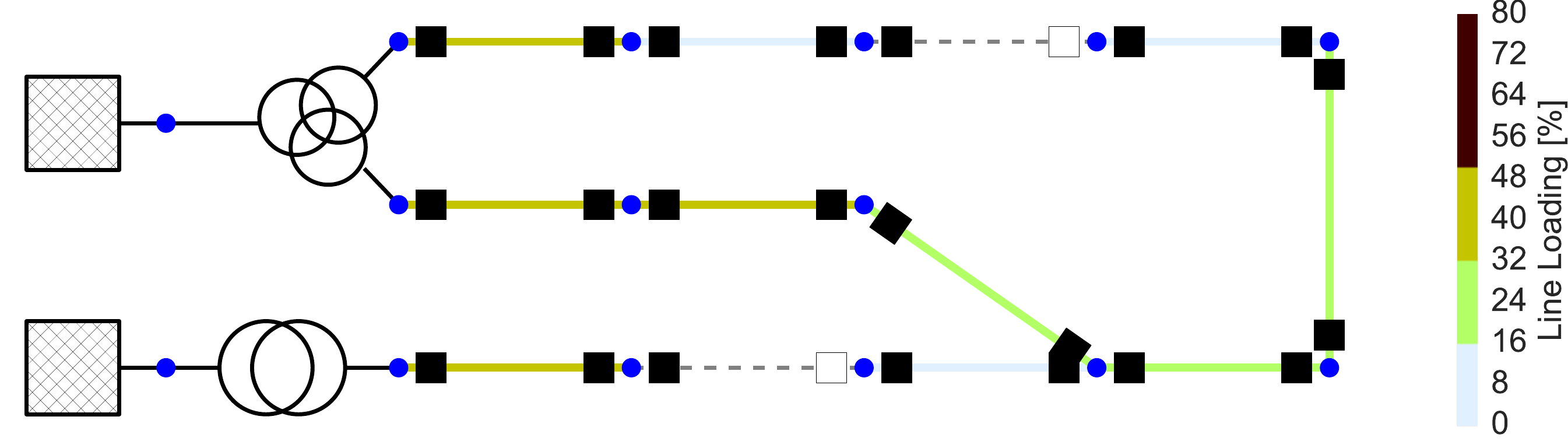} \label{fig:case_study:colormap1}}
	
	\subfloat[Grid state for peak load at 19:15 h]{\includegraphics[width=1.\textwidth]{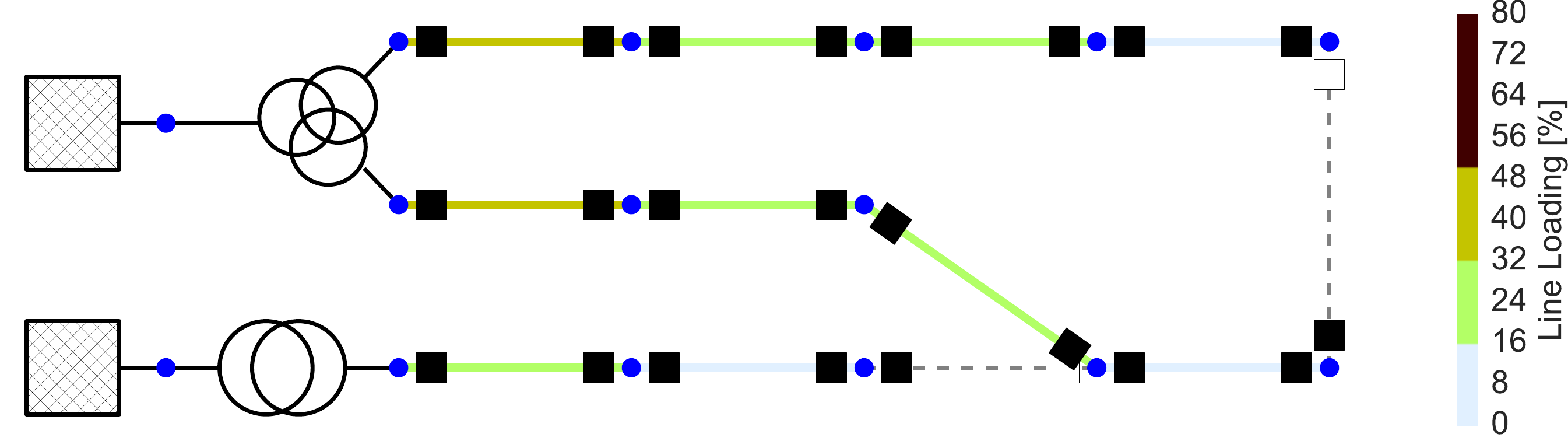} \label{fig:case_study:colormap2}}
  \caption{Grid state in two exemplary time steps \label{fig:case_study:colormaps}}
\end{figure}

The results of the time-series simulation with the approach outlined above is shown in Figure~\ref{fig:case_study:result}. It can be seen that the line and transformer loadings are always below \unit[50]{\%}, and the bus voltages always stay between \unit[0.97]{pu} and \unit[1.03]{pu}. The tap positions are adapted throughout the day in both the two-winding as well as the three-winding transformer between the positions -2 and +2. It can be seen, that the changes in the tap changer positions correlate with the changes in the bus voltages. The minimum fault current is  above \unit[1.1]{kA} at all times as specified. The grid is reconfigured throughout the day for an optimal state of operation with 9 different switching states. The changes in switching state also  coincide with changes in the fault currents, which are predominantly defined by the switching position. Figure \ref{fig:case_study:colormaps} shows two exemplary time steps for the peak generation situation at \unit[03:30]{h} and the peak load situation at \unit[19:15]{h}. It can be seen that different feeder configurations are applied to allow an optimal distribution of the power flows in each time step. A total of 9732 power flows are carried out for the time-series simulation including controller loops for tap-changer adaptations. This results in a total run time of \unit[7]{min} on a modern business laptop. However, since the time-steps are independent in a quasi-static simulation, the time-series simulation can also be parallelized. For the given case study, parallelized execution on a 48-core system reduced execution to under \unit[15]{s}. 

\subsection{Implementation in pandapower}
The results show that the algorithm implemented with \texttt{pandapower} successfully simulates an active grid operation with reconfiguration and tap changer control. Topological restrictions, line loadings, bus voltages, fault currents and active power losses are considered. The case study is focused on grid operation with radial operation, for which the \texttt{pandapower} switch model is necessary. It also makes use of the detailed transformer models, which include two-and three-winding transformers with tap changer functionality. The case study grid can be defined with all relevant parameters with 32 lines of code, which shows the user friendliness of the pandapower API for grid definition. The \texttt{pandapower} plotting module is used to create the grid plots shown in Figure~\ref{fig:case_study:grid} as well as in Figure~\ref{fig:case_study:colormaps}.

For the time-series simulation, the \texttt{pandapower} topology package is used to check for radiality as well as to identify the voltage control areas of the transformers depending on the switching state. The short-circuit module is used to check fault current magnitudes, and power flow calculations are used to analyze bus voltages as well as line and transformer loadings. The entire simulation can be implemented based on \texttt{pandapower} with a total of only 71 lines of code.

This example demonstrates that \texttt{pandapower} provides a wide range of predefined functions, which make the tool user friendly, while still providing the flexibility to implement customized applications. The fact that Python is  available as open source software and provides cross-platform compatibility, allows convenient deployment on multi-core systems, which can greatly reduce run time of computationally complex tasks. 

\subsection{Comparison to other tools}
The presented case study could not be reproduced easily with any other available tool. Most other open source tools do not allow to model the grid in such detail, especially with regard to three-winding transformers and tie-line switches (see Table~\ref{tab:elements}). No other open source tool does provide the graph search capability that is needed to check for radiality as well as to identify transformer voltage control areas. Fault current calculation in compliance with IEC 60909 that are used to check for minimal fault currents is also not available in any other open source tool. While workarounds could be implemented to generally model the desired behavior with other tools, it would be much more time-consuming than using the capabilities for grid modeling and analysis which are provided by \texttt{pandapower} out of the box. The relevant \texttt{pandapower} functions and models are also thoroughly tested, which makes the usage less prone to errors than custom implementations. Commercial tools would provide the necessary grid models and analysis functions to implement this case study. However, a specific use case can only be implemented if it is allowed by the API. For example, most commercial tools have internal graph search capabilities, but the user does not necessarily have direct access to those features. This limits the modeling flexibility compared to open source tools. If the provided example of a transformer control with flexible switch configurations is not explicitly provided through the API of the commercial tool, it would be much more difficult to implement than with an open source tool. In addition, commercial tools require a license, which limits the usage in general, and especially with regards to the deployment on multi-core clusters. 

\section{Conclusion}\label{sec:Conclusion}
This paper introduced the open source power systems analysis tool \texttt{pandapower}, which is aimed at automation of static and quasi-static analysis and optimization in balanced three-phase power systems. \texttt{pandapower} comes with static equivalent circuit models for electric elements that can be defined with common nameplate parameters. All electric power system models are thoroughly tested against commercial software tools and allow industry standard modeling of power systems. An ideal switch model as it is implemented in \texttt{pandapower}, to the best of our knowledge, has not been described anywhere in the literature or has been implemented in any other open source tool. Convenience functions for element definition as well as standard type libraries make it easy for the user to automate the creation of networks, which allows a convenient import of network data from different formats. The tabular data structure allows to easily access and analyze input and output parameters. For power flow studies, \texttt{pandapower} uses a jit-compiled, accelerated version of the \textsc{pypower} Newton-Raphson algorithm as the default solver. Additional enhancements have been made to the \textsc{pypower} solver, such as the capabilities to model constant current loads, asymmetric impedances or grids with multiple slacks. A connectivity check allows to calculate power flow with unsupplied areas. In addition to power flow and optimal power flow calculations, \texttt{pandapower} is the first open source power systems analysis tool to include a state estimation as well as short circuit calculation according to IEC 60909. Furthermore, an interface for topological searches allows the user to carry out customized graph searches on the electric network. \texttt{pandapower} also includes a plotting library for the convenient visualization of power systems. This exclusive network analysis functionality, in combination with the extensive model library, makes \texttt{pandapower} a valuable and innovative contribution to existing open source tools. An exemplary case study that showcases the unique capabilities of \texttt{pandapower} was presented in this paper. The case study especially highlights the easy analysis of time series data from quasi-static simulations and possible multi-core calculation features without any additional licensing fees. The code for the presented case study is available on github \cite{casestudy}. The easy usage, as demonstrated by the case study, makes pandapower well suited for applications in scientific studies as well as for educational purposes.

\texttt{pandapower} is under continuous development on github~\cite{pandapower_github}, and an extensive suite of unit and regression tests ensures the soundness and integrity of the implementation. A detailed documentation and interactive tutorials make \texttt{pandapower} easy to learn. Additional features such as an unbalanced power flow implementation, unbalanced short circuit calculations and a graphical user interface are planned to be added in the future.

\section*{Acknowledgement}
The development of \texttt{pandapower} was supported by the German Federal Ministry for Economic Affairs and Energy and the Projekttr\"ager J\"ulich GmbH (PTJ) within the framework of the projects \textit{Smart Grid Models} (FKZ: 0325616), \textit{OpSimEval} (FKZ: 0325782A). We acknowledge the feedback and contributions of all users that have helped to improve \texttt{pandapower}. We especially thank Jakov Krstulovi\'{c} Opara for contributing the ZIP load model, backward/forward sweep power flow and plotly interface.

\ifCLASSOPTIONcaptionsoff
  \newpage
\fi

\bibliographystyle{IEEEtran}
\bibliography{bibtex}
\vspace{-21 mm}

\begin{IEEEbiography}[{\includegraphics[width=1in,height=1.25in,clip,keepaspectratio]{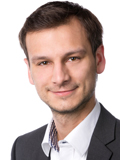}}]{Leon Thurner} received a B.Sc. degree in electrical engineering and business administration at the Technical University Kaiserslautern, Germany in 2011 and a M.Sc. degree in renewable energies and energy efficiency from the University of Kassel, Germany, in 2013. He is now a Ph.D. student at the Department of Energy Management and Power System Operation, University of Kassel, Germany. His main field of interest is automated network planning in distribution systems.
\end{IEEEbiography}
\vspace{-14 mm}
\begin{IEEEbiography}[{\includegraphics[width=1in,height=1.25in,clip,keepaspectratio]{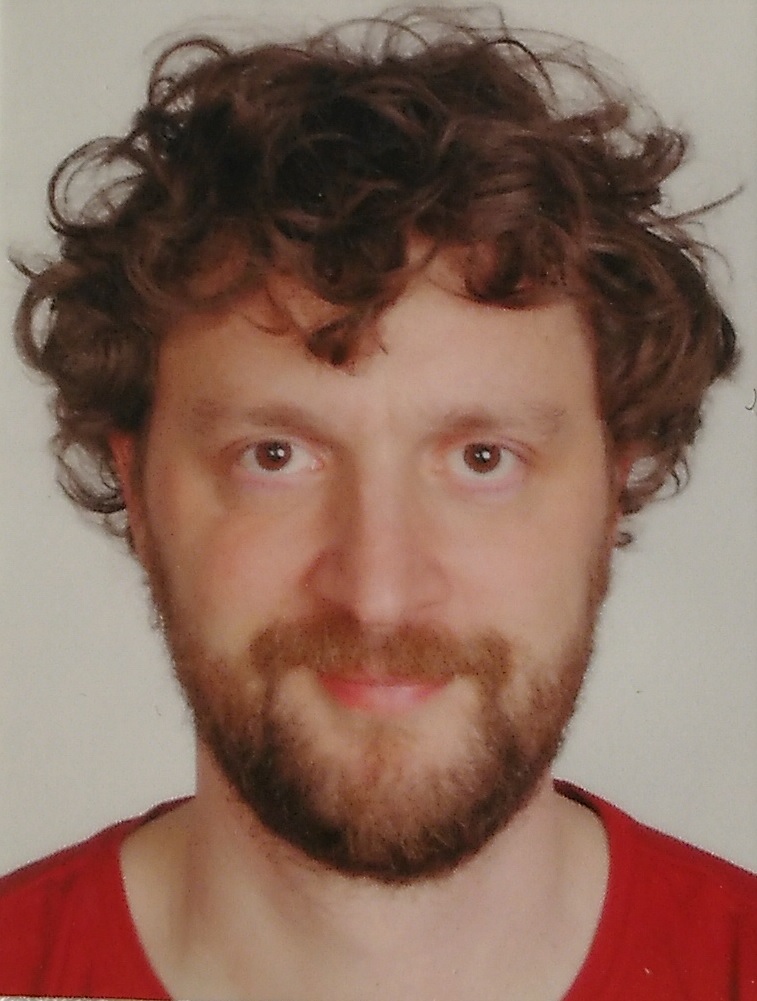}}]{Alexander Scheidler} received the Diploma and Ph.D. degrees in computer science from the University of Leipzig, Leipzig, Germany, in 2005 and 2010, respectively. He was a Post-Doctoral Fellow with the IRIDIA, Université Libre de Bruxelles, Brussels, Belgium, from 2010 to 2012 and has been with the Fraunhofer IEE, Kassel, Germany, since 2012. His current research interests include swarm intelligence, optimization methods for distribution grid planning and HPC in power system research.
\end{IEEEbiography}
\vspace{-12 mm}
\begin{IEEEbiography}[{\includegraphics[width=1in,height=1.25in,clip,keepaspectratio]{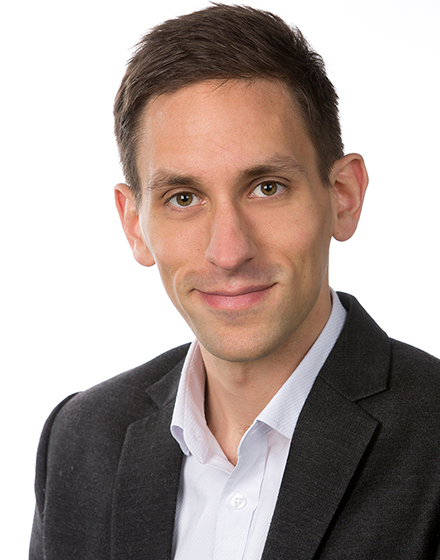}}]{Florian Sch\"afer} received  his  B.Sc.  and  M.Sc.  degrees  in Electrical Engineering, Information Technology and Computer Engineering  from the RWTH Aachen University, Germany, in 2013 and 2016, respectively. He is now a Ph.D. student at the Department of Energy Management and Power System Operation, University of Kassel, Germany. His main fields of interest are time series based power system planning strategies.
\end{IEEEbiography}
\vspace{-11 mm}
\begin{IEEEbiography}[{\includegraphics[width=1in,height=1.25in,clip,keepaspectratio]{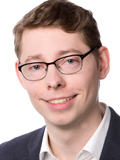}}]{Jan-Hendrik Menke} is pursuing the Ph.D. degree at the Department of Energy Management and Power System Operation, University of Kassel. He  received  his  B.Sc.  and  M.Sc.  degrees  in  electrical  engineering  from  the TU Dortmund University, Germany, in 2012 and 2014, respectively. He is interested in distribution system state estimation and machine learning in general.
\end{IEEEbiography}
\vspace{-10 mm}
\begin{IEEEbiography}[{\includegraphics[width=1in,height=1.25in,clip,keepaspectratio]{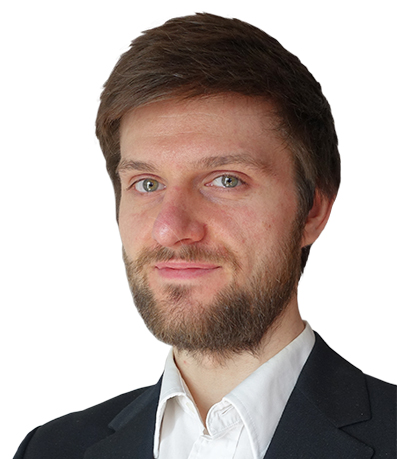}}]{Julian Dollichon} received  his  B.Sc. degree in Computer Science from the University of Kassel, Germany in 2015. Since 2014 he has been working as a student assistant at the Fraunhofer IEE in Kassel, Germany. He is currently pursuing his M.Sc. degrees in Computer Science at the University of Kassel. 
\end{IEEEbiography}
\vspace{-11 mm}
\begin{IEEEbiography}[{\includegraphics[width=1in,height=1.25in,clip,keepaspectratio]{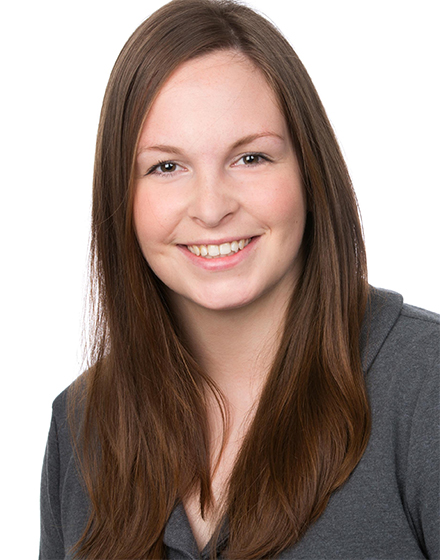}}]{Friederike Meier} received the B.Sc. and M.Sc. degrees in electrical engineering from the University of Kassel, Germany in 2013 and 2015. In 2015, she joined the Department of Energy Management and Power System Operation of the University of Kassel, Germany where she is pursuing her Ph.D. Since 2017, she has been a researcher at the Fraunhofer IEE in Kassel, Germany. Her current research interests are the techno-economic assessment of active power curtailment of renewables in grid planning.
\end{IEEEbiography}
\vspace{-12 mm}
\begin{IEEEbiography}[{\includegraphics[width=1in,height=1.25in,clip,keepaspectratio]{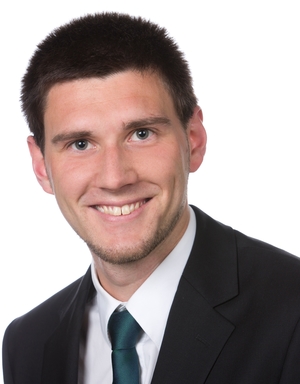}}]{Steffen Meinecke} is a Ph.D. student at the Department of Energy Management and Power System Operation of the University of Kassel, Germany. He received the B.Sc. and M.Sc. degrees in electrical engineering from the University of Kassel, Germany in 2013 and 2016. His research field is developing and analyzing appropriate benchmark grid datasets.
\end{IEEEbiography}
\vspace{-10 mm}
\begin{IEEEbiography}[{\includegraphics[width=1in,height=1.25in,clip,keepaspectratio]{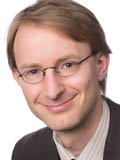}}]{Martin Braun} received Diploma degrees in electrical engineering as well as in technically oriented business administration from the University of Stuttgart, Germany in 2005 and a Ph.D. in engineering from  the  University  of  Kassel,  Germany,  in  2008.  He  is  now  Professor  of the  Department  of  Energy  Management  and  Power  System  Operation  at  the University of Kassel, Germany and director of the business field grid planning and grid operation at the Fraunhofer IEE, Kassel, Germany.
\end{IEEEbiography}

%
%
%

\end{document}